\documentclass[aip,
 amsmath,amssymb,
reprint,%
]{revtex4-1}

\usepackage{graphicx}
\usepackage{dcolumn}
\usepackage{bm}

\usepackage[utf8]{inputenc}
\usepackage[T1]{fontenc}
\usepackage{mathptmx}
\usepackage{etoolbox}

\makeatletter
\def\@email#1#2{%
 \endgroup
 \patchcmd{\titleblock@produce}
  {\frontmatter@RRAPformat}
  {\frontmatter@RRAPformat{\produce@RRAP{*#1\href{mailto:#2}{#2}}}\frontmatter@RRAPformat}
  {}{}
}%
\makeatother
\usepackage[colorlinks=true,linkcolor=blue,citecolor=blue,urlcolor=blue]{hyperref}

\begin{document}

\preprint{AIP/123-QED}

\title{Investigating the Interplay between Spin-Polarization and Magnetic Damping in $\mathrm{Co}_{x}\mathrm{Fe}_{80-x}\mathrm{B}_{20}$ for Magnonics Applications}
\author{Lorenzo Gnoatto}
\affiliation{Department of Applied Physics and Science Education, Eindhoven University of Technology, P.O. BOX 5132, 5600 MB Eindhoven, The Netherlands}
    \email{l.g.gnoatto@tue.nl}
\author{Thomas Molier}
\affiliation{Department of Applied Physics and Science Education, Eindhoven University of Technology, P.O. BOX 5132, 5600 MB Eindhoven, The Netherlands}

\author{Jelte J. Lamberts}
\affiliation{Department of Applied Physics and Science Education, Eindhoven University of Technology, P.O. BOX 5132, 5600 MB Eindhoven, The Netherlands}

\author{Artim L. Bassant}
\affiliation{Institute for Theoretical Physics, Utrecht University, Princetonplein 5, 3584 CC Utrecht, The Netherlands}

\author{Casper F. Schippers}
\affiliation{Department of Applied Physics and Science Education, Eindhoven University of Technology, P.O. BOX 5132, 5600 MB Eindhoven, The Netherlands}

\author{Rembert A. Duine}
\affiliation{Department of Applied Physics and Science Education, Eindhoven University of Technology, P.O. BOX 5132, 5600 MB Eindhoven, The Netherlands}
\affiliation{Institute for Theoretical Physics, Utrecht University, Princetonplein 5, 3584 CC Utrecht, The Netherlands}

\author{Reinoud Lavrijsen}
\affiliation{Department of Applied Physics and Science Education, Eindhoven University of Technology, P.O. BOX 5132, 5600 MB Eindhoven, The Netherlands}

\date{\today}

\begin{abstract}
For magnonics and spintronics applications, the spin polarization ($P$) of a transport current and the magnetic damping ($\alpha$) play a crucial role, e.g. for magnetization dynamics and magnetization switching applications. In particular, $P$ in a glassy (amorphous) 3d transition ferromagnet such as CoFeB and $\alpha$ are both strongly affected by $s-d$ scattering mechanisms. Hence, a correlation can be expected which is a priori difficult to predict. In this work, $P$ and $\alpha$ are measured using current-induced Doppler shifts using propagating spin-wave spectroscopy and broadband ferromagnetic resonance techniques in blanket films and current-carrying $Co_{\rm x}Fe_{\rm {80-x}}B_{\rm 20}$ alloy microstrips. The measured $P$ ranges from 0.18 $\pm$ 0.05 to 0.39 $\pm$ 0.05 and $\alpha$ ranges from $(4.0\pm 0.2)\cdot10^{-3}$ to $(9.7\pm 0.6)\cdot10^{-3}$. We find that for increasing $P$ a systematic drop in $\alpha$ is observed, indicating an interplay between magnetic damping and the spin polarization of the transport current which suggests that interband scattering dominates in $Co_{\rm x}Fe_{\rm {80-x}}B_{\rm 20}$. Our results may guide future experiments, theory, and applications in advancing spintronics and metal magnonics.
\end{abstract}

\maketitle


Metal-based spintronics leverages the spin polarization $P$ of itinerant electrons to spatially transfer angular momentum, allowing control over magnetization dynamics and magnetization switching\cite{manipatruni2018beyond}. Beyond conventional spintronics and magnonics—the manipulation of spin waves for logic operations and computations\cite{mahmoud2020introduction}—has emerged as a complementary field. Spin-polarized currents play a crucial role in magnonic systems, allowing current-induced spin-wave frequency shifts\cite{vlaminck2008current,haidar2013thickness} and modulation of amplitude\cite{gladii2016spin,an2014control}. 
At the heart of both spintronics and magnonics lies the intricate relationship between spin polarization $P$, magnetic damping $\alpha$, and the electronic density of states (DOS) at the Fermi level. Spin transfer torques, driven by spin-polarized currents, induce phenomena such as spin wave Doppler shifts\cite{haidar2013thickness}, which provides access to the magnitude of $P$. Simultaneously in 3d transition metals, $s-d$ scattering, while contributing to high damping by dissipating spin wave (SW) energy, is also responsible for generating the spin-polarized current, as spin-up and spin-down carriers experience different scattering efficiencies in ferromagnets\cite{zhu2011enhanced,thomas2011impact}. 

Recent studies have independently examined the influence of the density of states (DOS) at the Fermi level on spin polarization $P$ and magnetic damping $\alpha$. In CoFeB alloys, a strong correlation between $P$ and alloy composition has been attributed to $s-d$ hybridization and compositional transitions in the DOS \cite{paluskar2009correlation}. However, in that approach, $P$ is measured via spin-polarized tunneling into a superconductor at a temperature of approximately 250 mK. As a result, the method probes the spin polarization of interface states at the insulating barrier rather than the bulk polarization, and the measurements are conducted far from room temperature. Similarly, magnetic damping has been shown to be minimized in CoFe alloys in specific compositions where the DOS at the Fermi level reduces spin-orbit coupling, offering valuable insights into the mechanisms underlying energy dissipation \cite{schoen2016ultra}. Based on this, theoretical studies have demonstrated that $\alpha$ and $P$ are intrinsically connected through spin-flip scattering processes, the electronic band structure, and spin-orbit coupling, which collectively govern both energy dissipation and spin transport in transition metal alloys \cite{starikov2010unified}. In particular, spin-orbit coupling and disorder are critical factors that influence damping, with interband scattering mechanisms playing a dominant role in the observed correlation between $\alpha$ and $P$. Despite these theoretical insights, an experimental direct investigation of the interaction between $P$ and $\alpha$—both strongly influenced by the electronic band structure—remains unexplored.

In this work, our aim is to address this knowledge gap by exploring the relationship between $P$ and $\alpha$ in glassy (amorphous) CoFeB alloys using propagating spin wave spectroscopy\cite{lucassen2019optimizing} (PSWS) that gives direct access to both parameters. These alloys are of significant interest for spintronics applications such as magnetic tunnel junctions\cite{ikeda2010perpendicular}, spin valves\cite{morgunov2017ferromagnetic}, and spin torque devices\cite{zahedinejad2018cmos}. Moreover, they are easy to sputter with high saturation magnetization $(M_{\rm s})$ needed for inductive methods, and lack crystalline structural transitions. By systematically characterizing the magnetic parameters of unpatterned and patterned films, we investigate how composition influences $P$ and $\alpha$ and discuss the possible interplay between these properties.
\begin{figure}
    \centering
    \includegraphics[width=1\linewidth]{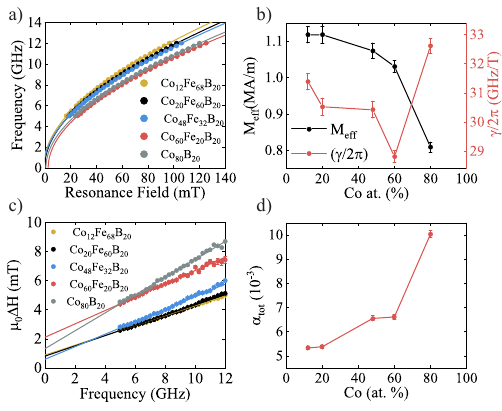}
    \caption{\label{main1} a) Frequency as a function of resonance field for the $Co_{\rm x}Fe_{\rm {80-x}}B_{\rm 20}$ series. b) extracted $M_{\rm eff}$ (left y-axis) and $\gamma/2\pi$ (right y-axis) as a function of composition. c) Linewidth as a function of excitation frequency and associated damping in d). }
\end{figure}

\begin{figure}
    \centering
    \includegraphics[width=1\linewidth]{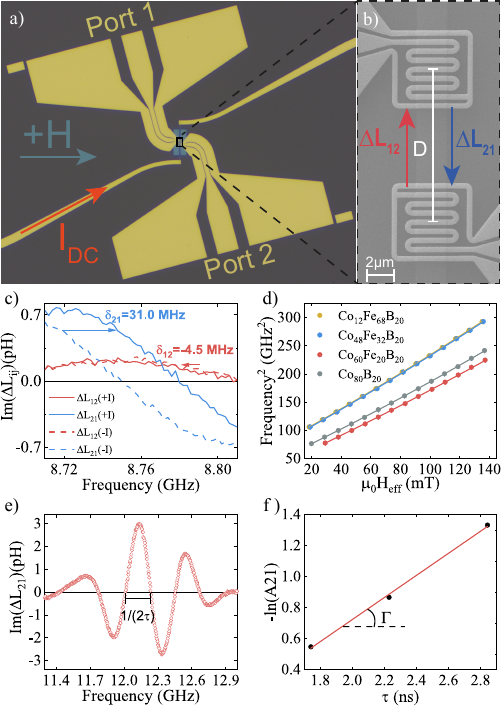}
    \caption{\label{main2} a) Optical microscope image of the PSWS device. b) Scanning electron microscope image of the spin wave antennas for a device with $D=9.57 \mu \mathrm{m}$. c) Mutual inductance spectra for a permalloy sample under a current of $|I|=\mathrm{8 \ mA}$ and external magnetic field of +60 mT. d) Dependence on the effective magnetic field of the resonance frequency for the $\mathrm{Co_{x}Fe_{ {80-x}}B_{20}}$ series, solid line is the fit from Eq.~\ref{eq:frequency_squared}. e) Measured mutual inductance for the composition $\mathrm{Co_{48}Fe_{32}B_{20}}$ with $D=9.57\ \mu \mathrm{m}$ and external applied field of $\mu_{\rm 0} H_{\rm ext}=$ 60 mT. f) Dependence of the logarithm of $A_{\rm 21}$ on $\tau$.}
\end{figure}

A series of $\mathrm{Ta(4)/Co_{x}Fe_{80-x}{B}_{20}(20)/Ta(4)}$ (numbers in parentheses are thicknesses in nanometers) with $x=\mathrm{(12, 20, 48, 60, 80)}$ thin film has been prepared via DC magnetron sputtering; see Suppl. Mat. for details. The bottom Ta layer improves adhesion, and the top layer prevents oxidation. Films used for PSWS and ferromagnetic resonance (FMR) were deposited simultaneously to ensure identical starting parameters. In-plane FMR spectra were measured using a flip-chip method on a coplanar waveguide at fixed frequency while sweeping the external magnetic field. Values of $M_{\text{eff}}$ and $\gamma$ were obtained by fitting to the Kittel relation, as shown in Fig.~\ref{main1}(a) (see Suppl. Mat. for details). The variation of $M_{\text{eff}}$ with the atomic percentage of Co in the alloy, as shown in Fig. 1(b), exhibits Slater-Pauling behavior, reaching a maximum at $x = 20$, consistent with previous reports\cite{paluskar2009correlation}. The extracted values for $\gamma$ per composition are shown in Fig.~\ref{main1}(b) and will be used later to extract $\alpha_{\text{PSWS}}$  and $P$ from the PSWS signals. In Fig.~\ref{main1}(c) the full width at half maximum allows us to extract $\alpha_{\text{FMR}}$ of the blanket samples showing an increase with increasing Co content $(x)$, this behavior has been reported for CoFe~\cite{schoen2016ultra} alloys. 
Fig. \ref{main2}(a) shows a wide field optical microscopy image of our PSWS measurement device and in \ref{main2}(b) a scanning electron micrograph of the center region is shown. It consists of a magnetic microstrip with a width of 2 $\mu \mathrm{m}$ fabricated by negative resist-based electron beam lithography (EBL) and argon ion milling. The magnetic strip is then electrically isolated from the antennas with a 50 nm thick layer of MgO, deposited using e-beam evaporation via a lift-off-based EBL step (see the light blue region in Fig. \ref{main2}(a)). The antennas and electrical contacts to the ferromagnetic strip are then defined using a third lift-off EBL step and e-beam evaporation of Ti(10)/Au(100). The antennas are optimized~\cite{lucassen2019optimizing} to excite spin waves with a wavenumber of 5.4 $\mu \mathrm{m}^{-1}$. The distances from center to center $D$ of the antennas are varied between 7.57, 9.57 and 11.57 $\mu \mathrm{m}$ allowing us to extract $\alpha_{PSWS}$ from the PSWS signals \cite{gladii2017spin} in addition to the values of $\alpha_{FMR}$ obtained from the FMR method.

\begin{figure}
    \centering
    \includegraphics[width=1\linewidth]{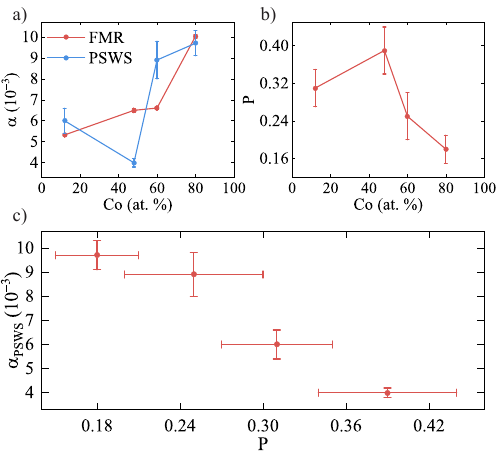}
    \caption{\label{main3} a) Magnetic damping as a function of composition from FMR and PSWS. b) Spin polarization as a function of alloy composition. c) Extracted $\alpha_{\text{PSWS}}$ as a function of $P$.}
\end{figure}

$P$ and $\alpha_{\rm PSWS}$ are measured using PSWS. To extract $\alpha_{\rm PSWS}$ we use the relation between the spin wave relaxation rate and $\alpha$ for in-plane magnetized films~\cite{gladii2017spin}:
\begin{align}
\Gamma &= \alpha_{\rm PSWS}\left(\omega_{\rm 0}+\frac{\omega_{\rm M}}{2}\right), \label{eq:rel_rate}
\end{align}
where $\omega_{\rm 0}=\gamma\mu_{\rm 0}H_{\rm eff}$ with $H_{\rm eff}=H_{\rm ext}+H_{\rm d}$ is the effective field taking into account the external applied field $H_{\rm ext}$ and demagnetizing field $H_{\rm d}$ typical of rectangular geometries\cite{bracher2017creation,joseph1965demagnetizing} and $\omega_{\rm M}=\gamma\mu_{\rm 0}M_{\rm s}$, see Suppl. Mat. for further details. Magnetic parameters such as $M_{\rm s}$ and $M_{\rm eff}$ have been calculated from fitting the center PSWS signal as a function of the effective magnetic field using\cite{gladii2017spin}: 
\begin{align}
f^2 &= \left( \frac{\mu_{\rm 0} \gamma}{2\pi} \right)^2 
\Bigg\{ (H_{\rm {eff}}^2 + H_{\rm eff}M_{\rm eff} \notag \\
&\quad + \frac{M_{\rm s} M_{\rm eff}}{4} \left[ 1 - \exp(-2k t) \right] \Bigg\}, \label{eq:frequency_squared}
\end{align}
where $\gamma$ is the gyromagnetic ratio, $M_{\rm eff}=M_{\rm s}-H_{\rm u}$  the effective magnetization that takes into account a possible small out-of-plane anisotropy $H_{\rm u}$, and $t$ is the nominal film thickness. In this fitting the $\gamma$ extracted from the FMR data shown in Fig.~\ref{main1}(b) is used. To evaluate the spin wave relaxation rate we define a normalized transmitted amplitude as $A_{\mathrm {21}}=|\mathrm{Im}(\Delta L_{\rm 21}|^{\rm max})/\sqrt{\mathrm{Im}(|\Delta L_{\rm 11}|^{\rm max})\mathrm{Im}(\Delta L_{\rm 22}|^{\rm max})}$ where $|\mathrm{Im}(\Delta L_{\mathrm {11}})|$ and $|\mathrm{Im}(\Delta L_{\mathrm {22}})|$ are the absolute value of the imaginary part of the self inductance (reflected signal) and $|\mathrm{Im(\Delta L_{21}})|$ the mutual inductance (transmitted signal)\cite{vlaminck2010spin}. The amplitude of the spin wave propagating over a distance $D$ decays exponentially as $A_{\rm 21}=\mathrm{exp}[-D/L_{\mathrm {att}}]$ where $L_{\mathrm {att}}$ is the attenuation length, in our work we consider $D$ to be the distance between the center position of the antennas~\cite{chang2014phase}.
In Fig.~\ref{main3}(e) we evaluate the group delay time $(\tau)$ for each distance $D$ and we plot $-\mathrm{ln}(A_{\rm 21})$ versus group delay time, Fig.~\ref{main2}(f). The slope is then proportional to the spin wave relaxation rate~\cite{gladii2017spin} $\Gamma$ in rad/s from which we calculate $\alpha_{\rm PSWS}$ using Eq.~\ref{eq:rel_rate}, which is shown in Fig. ~\ref{main3}(a). Within measurement accuracy we obtain similar values of $\alpha_{\rm FMR}$ and $\alpha_{\rm PSWS}$ indicating no significant increase of the damping between the blanket (FMR) and patterned CoFeB films (PSWS). 
To extract $P$ we have measured the shift in the PSWS signals (Doppler shift) as a function of applied current while in the magnetostatic surface wave mode (Damon Eshbach). Ignoring non-reciprocities the pure spin-transfer-torque induced Doppler shift~\cite{haidar2013thickness,zhu2011enhanced} is defined as:
\begin{align}
\Delta f_\text{dop} &= \frac{\delta f_{\rm 12} - \delta f_{\rm 21}}{4} 
= -\frac{g \mu_\text{B} P}{4 \pi M_{\rm s} |e|} \frac{I_{\text{FM}}}{w\ t}k, \label{eq:doppler_frequency}
\end{align}
where $\delta f_{\rm 12}$ and $\delta f_{\rm 21}$ are the current induced frequency shift for oppositely propagating spin waves, $g$ is the g-factor related to $\gamma$ via $g=\gamma\  \hbar /\mu_{B} $, $\mu_{\text{B}}$ the Bohr magneton, $M_{\rm s}$ the saturation magnetization, $e$ the electron charge, $k$ the spin-wave wavevector, $w$ is the width of the ferromagnetic strip and $I_{\text{FM}}$ the current through the ferromagnet, see Suppl.Mat. for details. This relation indicates that $\Delta f_\text{dop} \propto I_{\text{FM}}$, hence, from extracting $\Delta f_\text{dop}$ as a function of $I_{\text{FM}}$, $P$ can be determined. However, in the magnetostatic spin wave mode, other sources of non-reciprocal propagation e.g. Oersted fields under current need to be taken into account in addition to reciprocal effects due to e.g. Joule heating. We compensate for this using the methodology developed by Haidar \textit{et al.} \cite{haidar2013thickness,haidar2012role}. We have validated our method by fabricating an similar device as Haidar \textit{et al.};  $\mathrm{Ta}(4)/\mathrm{Py}(20)/\mathrm{Ta}(4)$, measured the PSWS signals under current (shown in Fig.~\ref{main2}(c)) and extracted $P=0.67\pm0.08$. This value is fully consistent with the earlier reports of $P$ in Permalloy (e.g. $~0.63\pm0.4$ as reported by Haidar \textit{et al.}\cite{haidar2013thickness}) confirming our method is sound. In Fig 2S (b-f) in the Suppl. Mat. we show the current induced frequency shift for all the $\mathrm{Co}_{x}\mathrm{Fe}_{80-x}\mathrm{B}_{20}$ series and permalloy. Finally, we plot $\alpha_{\rm PSWS}$ as a function of $P$ in Fig.~\ref{main3}(c), where $\alpha_{\rm PSWS}$ decreases systematically with greater $P$. The observed inverse correlation between spin polarization $P$ and magnetic damping $\alpha$ in our $\mathrm{Co}_{x}\mathrm{Fe}_{80-x}\mathrm{B}_{20}$ alloys can be linked to spin-flip scattering processes and spin-orbit coupling effects, as discussed by Starikov \textit{et al.}\cite{starikov2010unified}, where Gilbert damping is shown to decrease with reduced intraband scattering contributions. However, to fully correlate our results with calculations the thermal lattice and spin disorder need to be taken into account. 

In summary, we have measured the spin polarization $P$ and damping $\alpha_{\mathrm{PSWS}}$ in different alloy compositions $\mathrm{Ta}(4)/\mathrm{Co}_{x}\mathrm{Fe}_{80-x}\mathrm{B}_{20}/\mathrm{Ta}(4)$ using PSWS and FMR techniques. Spin polarization increases systematically with lower damping values, highlighting a strong inverse relationship between these parameters. Furthermore, we find that PSWS is the most suitable technique for this type of study, as it uniquely enables probing of spin polarization $P$ within most of the material. This capability is particularly advantageous for thick films (>20 nm), such as those examined in this study, providing insights in the bulk mechanism compared to surface-sensitive methods.

\section*{Acknowledgements} We thank Paul Kelly for discussions of our data. This publication is part of the project "Black holes on a chip" with file number OCENW.KLEIN.502 of the research programme NWO Open Competitie ENW - KLEIN which is financed by the Dutch Research Council (NWO).

\section*{author contributions}
\textbf{Lorenzo Gnoatto:} Conceptualization (equal); Data curation (lead); Investigation (lead); Methodology (lead); writing - original draft (lead); Writing - review \& editing (lead). \textbf{Thomas Molier:} Conceptualization (equal); Data curation (equal); Investigation (equal); Methodology (equal); Writing - review \& editing (supporting); \textbf{Jelte J. Lamberts:} Investigation (supporting); Writing - review \& editing (supporting). \textbf{Artim L. Bassant:} Conceptualization (supporting); Writing - review \& editing (supporting). \textbf{Casper F. Schippers:} Data curation (equal); Writing - review \& editing (supporting). \textbf{Rembert A. Duine:} Conceptualization (equal); Funding acquisition (equal); Supervision (supporting);  Validation (equal);  Writing –
original draft (supporting); Writing – review \& editing (equal); \textbf{Reinoud Lavrijsen:} Conceptualization (equal); Formal analysis (equal); Funding acquisition (lead); Investigation (equal); Methodology (equal); Supervision (lead); Validation (equal); Writing – original draft (equal); Writing – review \& editing (equal).

\section*{Data Availability Statement}

The data that support the findings of this study are available from the corresponding author upon reasonable request

\section{References}
\bibliography{Main_RL_ADAPT}
\end{document}


\preprint{AIP/123-QED}

\title{Supplementary material: Investigating the Interplay between Spin-Polarization and Magnetic Damping in $\mathrm{Co}_{x}\mathrm{Fe}_{80-x}\mathrm{B}_{20}$ for Magnonic Applications}
\author{Lorenzo Gnoatto}
\affiliation{Department of Applied Physics and Science Education, Eindhoven University of Technology, P.O. BOX 5132, 5600 MB Eindhoven, The Netherlands}
    \email{l.g.gnoatto@tue.nl}
\author{Thomas Molier}
\affiliation{Department of Applied Physics and Science Education, Eindhoven University of Technology, P.O. BOX 5132, 5600 MB Eindhoven, The Netherlands}

\author{Jelte J. Lamberts}
\affiliation{Department of Applied Physics and Science Education, Eindhoven University of Technology, P.O. BOX 5132, 5600 MB Eindhoven, The Netherlands}

\author{Artim L. Bassant}
\affiliation{Institute for Theoretical Physics, Utrecht University, Princetonplein 5, 3584 CC Utrecht, The Netherlands}

\author{Casper F. Schippers}
\affiliation{Department of Applied Physics and Science Education, Eindhoven University of Technology, P.O. BOX 5132, 5600 MB Eindhoven, The Netherlands}

\author{Rembert A. Duine}
\affiliation{Department of Applied Physics and Science Education, Eindhoven University of Technology, P.O. BOX 5132, 5600 MB Eindhoven, The Netherlands}
\affiliation{Institute for Theoretical Physics, Utrecht University, Princetonplein 5, 3584 CC Utrecht, The Netherlands}

\author{Reinoud Lavrijsen}
\affiliation{Department of Applied Physics and Science Education, Eindhoven University of Technology, P.O. BOX 5132, 5600 MB Eindhoven, The Netherlands}

\date{\today}

\maketitle
\section{DC Magnetron sputtering}
The samples characterized in this work were grown by DC magnetron sputtering on Si substrates with 100 nm of thermal oxide. The general stack structure is Ta(4)/FM(20)/Ta(4), where FM indicates the ferromagnetic active layer, and the numbers in parentheses indicate layer thicknesses in nanometers. The sputter deposition system isoperated at a typical base pressure $< 5 \cdot 10^{8}$ mBar, with other details of the deposition system described elsewhere\cite{kools2023aging}.

\section{\label{sec:level1}FMR}
\begin{figure}
    \centering
    \includegraphics[width=0.5\linewidth]{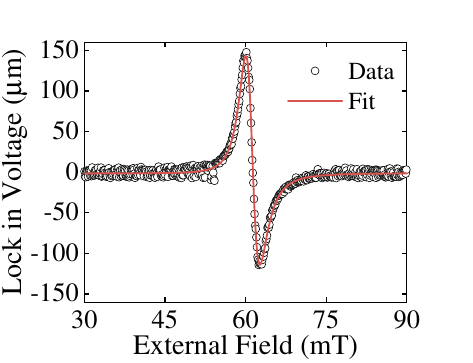}
    \caption{\label{fig1_SUP} Flip-Chip FMR absorption spectrum example for $Co_{48}Fe_{32}B_{20}$ at 9 GHz excitation and related fit to determine centre resonance and linewidth.}
\end{figure}

Using a flip-chip FMR setup with an AC-modulation coil configuration and a diode rectifier, we measured the power absorption as a function of the external magnetic field. The data were fitted using a superposition of symmetric and anti-symmetric Lorentzians, as shown in Fig.~\ref{fig1_SUP}. The field dependence of the extracted center frequency was then fitted using a modified Kittel relation\cite{beik2017broadband}.
\begin{align}
f &= \frac{\gamma}{2\pi} \sqrt{(B_\text{res} + B_\text{u})(B_\text{res} + B_\text{u} + \mu_0 M_\text{eff})} \label{eq:resonance_frequency},
\end{align}
where $\gamma$ is the gyromagnetic ratio, $B_{\rm res}$ is the resonance field, $B_{\rm u}$ takes into account a possible uniaxial anisotropy field which could be induced during sputter deposition. Finally, $M_{\rm eff}$ is the effective magnetization. 
The damping parameters were determined by fitting the extracted linewidth $\Delta B$ as a function of magnetic field:
\begin{align}
\Delta B &= \frac{4\pi}{\gamma} \alpha f + \Delta B_{\rm 0},\label{eq:delta_B},
\end{align}
where $\Delta B_{\mathrm 0}$ is the inhomogeneous broadening term and $f$ the frequency. This allows us to extract the damping for every CoFeB composition as plotted in the main text. 

\section{PSWS}
\subsection{Spin wave Doppler shift ($\Delta f_{\text{dop}}$) data analysis}

\begin{figure}
    \centering
    \includegraphics[width=1\linewidth]{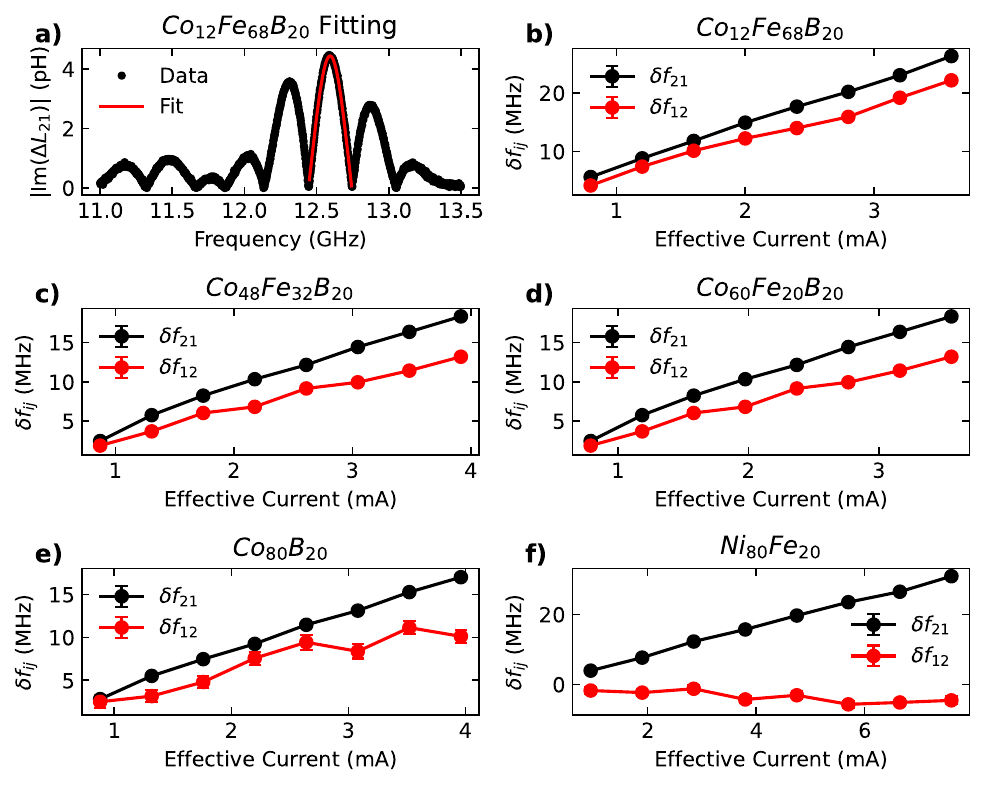}
    \caption{\label{fig2_SUP} a) Spectra showing the mutual inductance $\mathrm{Im}|\Delta L_{21}|$ and Gaussian fitting to determine the frequency shift. (b-e) Current induced frequency shift as a function of the effective current calculated via parallel resistor model for all the $\mathrm{Ta}(4)/\mathrm{Co}_{x}\mathrm{Fe}_{80-x}\mathrm{B}_{20}/\mathrm{Ta}(4)$ series. f) Same as (b-e) but for the Ta(4)/Py(20)/Ta(4) stack.  }
\end{figure}

In our data analysis we consider $(+k)$ to be spin waves (SWs) travelling from port 1 to port 2 ($\Delta L_\mathrm{21}$) and $(-k)$ SW's travelling from port 2 to port 1 ($\Delta L_\mathrm{12}$). In Fig. 2(c) of the main text, we observe that for $(+k)$ SWs (in blue) the $(+I)$ curve lies at higher frequency than for $(-I)$, $\delta f_\mathrm{21}=f_\mathrm{21}(+I)-f_\mathrm{21}(-I)=+31\ \text{MHz}$. Conversely, for $(-k)$ (in red), the $(+I)$ curves lies at lower frequencies than for $(-I)$, $\delta f_\mathrm{12}=f_\mathrm{12}(+I)-f_\mathrm{12}(-I)=-4.5$ MHz, this behaviour is expected for a positive value of spin polarization $P$, the full set of data are shown in Fig.~\ref{fig2_SUP}. From Fig.~\ref{fig2_SUP} one can recognize a different behaviour between the $\mathrm{Ta}(4)/\mathrm{Co}_{\rm x}\mathrm{Fe}_{\rm {80-x}}\mathrm{B}_{\rm 20}(20)/\mathrm{Ta}(4)$ series and the Ta(4)/Py(20)/Ta(4). This can be attributed to a dominating Oersted field effect in the CoFeB due to its high resistivity. This leads to a different current distribution in the stacks i.e. in the case of CoFeB more current will be shunted through the Ta layers. The measured shift is further augmented by the fact that CoFeB has lower $P$ than Py as we saw in the main text. To address these issues we use the method developed by Haidar et al.\cite{haidar2013thickness} which accounts for the inhomogeneous Oersted field distribution across the stack. To test our methodology we first reproduced their measurements on Ta(4)/Py(20)/Ta(4) and obtain a similar $P$ to their work using $\mathrm{Al}_2\mathrm{O}_3(\mathrm{21})/\mathrm{Py(20)}/\mathrm{Al}_2\mathrm{O}_3$, $0.67\pm0.8$ and $~0.63\pm0.4$, respectively after correcting for current shunting as shown in the next section.

\subsection{Parallel conductor model}
The current flowing in the magnetic layer is obtained using a parallel-conductor model that takes into account the resistivity ratio between the active layer and the underlying and capping Ta layers, similar to the approach in Zhu et al.~\cite{zhu2011enhanced}. We used in-line four-point sheet resistance measurements on all the deposited stacks and on a sample with 8 nm of Ta, which we assume represents the two 4 nm thick Ta layers, ignoring the CoFeB-Ta interface scattering. We justify this assumption by considering that the two interfaces contribute an equal amount of scattering in both the Ta and CoFeB layers, and hence the ratio of shunting is not changed. From this data, we calculate the corrected resistance for each layer (and later resistivity):
\begin{align}
\frac{1}{R_{\rm tot}} &= \frac{1}{R_{\rm FM}}+\frac{1}{R_{\rm Ta}} \label{eq:resistor},
\end{align}
where $R_{\rm tot}$ is the resistance of the full stack, $R_{\rm FM}$ is the resistance of the ferromagnet and $R_{\rm Ta}$ the resistance of the Tantalum layers. Under the assumption of uniform film thickness, the resistivity of the thin film can be defined as $\rho_{\rm sheet}= R_{\rm sheet}\ t$ where $R_{\rm sheet}$ is the measured sheet resistance and $t$ is the nominal film thickness. By knowing the film thicknesses of the layers from Eq.~\ref{eq:resistor} we calculate the resistivity of the ferromagnet using:
\begin{align}
\rho_{\mathrm{FM}} &= \frac{\rho_{\rm tot}\ \rho_{\rm Ta}\ t_{\rm FM}}{\rho_{\rm Ta}\ t_{\rm tot}\ -\ \rho_{\rm tot}\ t_{\rm Ta} } \label{eq:resistor2},
\end{align}
where $\rho_{\rm tot}$ and $\rho_{\rm Ta}$ are the measured resistivity of the total stack and 8 nm Ta layer, respectively, with $t_{\rm tot}$ and $t_{\rm Ta}$ their thicknesses. The ratios $I_{\rm tot}/I_{\rm FM}$ where $I_{\rm tot}$ is the applied current and $I_{\text{FM}}$ is the current flowing in the active layer gives the correction coefficients ($C_{\mathrm{corr}}$) for the applied current: 
\begin{align}
C_{\mathrm{corr}} &= \frac{ \rho_{\rm Ta}\ t_{\rm FM}}{\rho_{\rm Ta}\ t_{\rm FM}\ +\ \rho_{\rm FM}\ t_{\rm Ta} }, \label{eq:resistor3}
\end{align}
The calculated values are reported in Table ~\ref{tab:table1}, where we indeed observe that in the case of Py a 5\% correction is required, where for the CoFeB stacks up to 23\% needs to be corrected.

\begin{table}
\caption{\label{tab:table1}Current correction coefficients calculated to account different current distribution in the layered stacks. }
\resizebox{\textwidth}{!}{%
\begin{ruledtabular}
\begin{tabular}{l c}
Thin film stack & $C_{\mathrm{corr}}= I_{\mathrm{tot}}/I_{\mathrm{FM}}$  \\
\hline
$\mathrm{Ta(4)/{Co_{12}Fe_{68}B_{20}(20)/Ta(4)}}$ & 0.83 \\
$\mathrm{Ta(4)/Co_{48}Fe_{32}B_{20}(20)/Ta(4)}$ & 0.77 \\
$\mathrm{Ta(4)/Co_{60}Fe_{20}B_{20}(20)/Ta(4)}$ & 0.80 \\
$\mathrm{Ta(4)/Co_{80}B_{20}(20)/Ta(4)}$ & 0.85 \\
$\mathrm{Ta(4)/Py(20)/Ta(4)}$ & 0.95 \\
\end{tabular}
\end{ruledtabular}
}%
\end{table}

\subsection{Training procedure}
\begin{figure}
    \centering
    \includegraphics[width=1\linewidth]{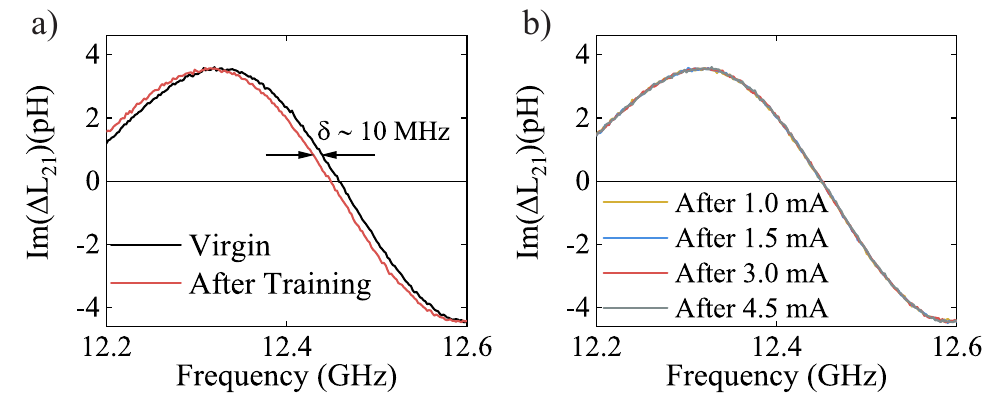}
    \caption{\label{fig3_SUP} a) Two subsequent measurement performed without applied current before (virgin) and after (after training) applying 5 mA for 45 minutes for $\mathrm{Ta}(4)/\mathrm{Co}_{\rm 48}\mathrm{Fe}_{\rm 32}\mathrm{B}_{\rm 20}(20)/\mathrm{Ta}(4)$. b) Four subsequent measurement performed with no applied current after applying the current presented in the legend: after the training routine  no change in the center frequency is observed. The same frequency shift is observed for $\Delta L_{12}$ and for all the other $\mathrm{Ta}(4)/\mathrm{Co}_{\rm x}\mathrm{Fe}_{\rm {80-x}}\mathrm{B}_{\rm 20}(20)/\mathrm{Ta}(4)$ compositions (not shown).}
\end{figure}

In our experiment we noticed that for the $\mathrm{Ta(4)/Co_{x}Fe_{80-x}{B}_{20}(20)/Ta(4)}$ series the magnetic and electric properties changed during subsequent Doppler shift measurements as shown in Fig.~\ref{fig3_SUP} (a), where a shift of \textasciitilde10 MHz is observed, under zero current, between the first measurement (virgin) on a device and after applying 5 mA for 45 minutes through the device. This effect is well known from e.g. CoFeB/MgO based magnetic tunnel junctions where the Boron diffuses out of the CoFeB during annealing at high temperatures (>225$^\circ$C)\cite{aleksandrov2016evolution,wang2016magnetic}. In our case for the highest applied current densities ($<10^{11}$ A / m$^2$), we expect the temperature rise to be below 50$^\circ$C at room temperature (i.e. max 80 $^\circ$C), and no crystallization can occur and we assume the system to remain amorphous at 20\% boron concentration\cite{kim2022control}.
To stabilize our samples before gathering a consistent data set, we empirically 'train' the samples by applying 5 mA for 45 minutes through each sample, which stabilizes the system, as can be seen in Fig.~\ref{fig3_SUP} (b) where the center frequency does not change after subsequent measurement if the sample has been trained. Note that this effect was not observed for Py, due to the lower resistivity and hence less Joule heating for similar current densities, and the absence of Boron. 

\section{Damping PSWS}
The magnetization along the short axis of the waveguide results in an inhomogeneous effective field across its width, we can approximate the effective field $H_\mathrm{eff}$ distribution inside the strip using\cite{bracher2017creation,joseph1965demagnetizing}:
\begin{align}
\mu_0 H_{\mathrm{eff}}(y) 
&= \mu_0 H_{\mathrm{ext}} - \frac{\mu_0 M_{\mathrm{s}}}{\pi} 
\left[ 
\arctan \left( \frac{t}{2(y - y_{\mathrm{0}}) + w} \right) 
- \arctan \left( \frac{t}{2(y - y_{\mathrm{0}}) - w} \right)
\right], \label{eq:demag}
\end{align}
\begin{figure}
    \centering
    \includegraphics[width=1\linewidth]{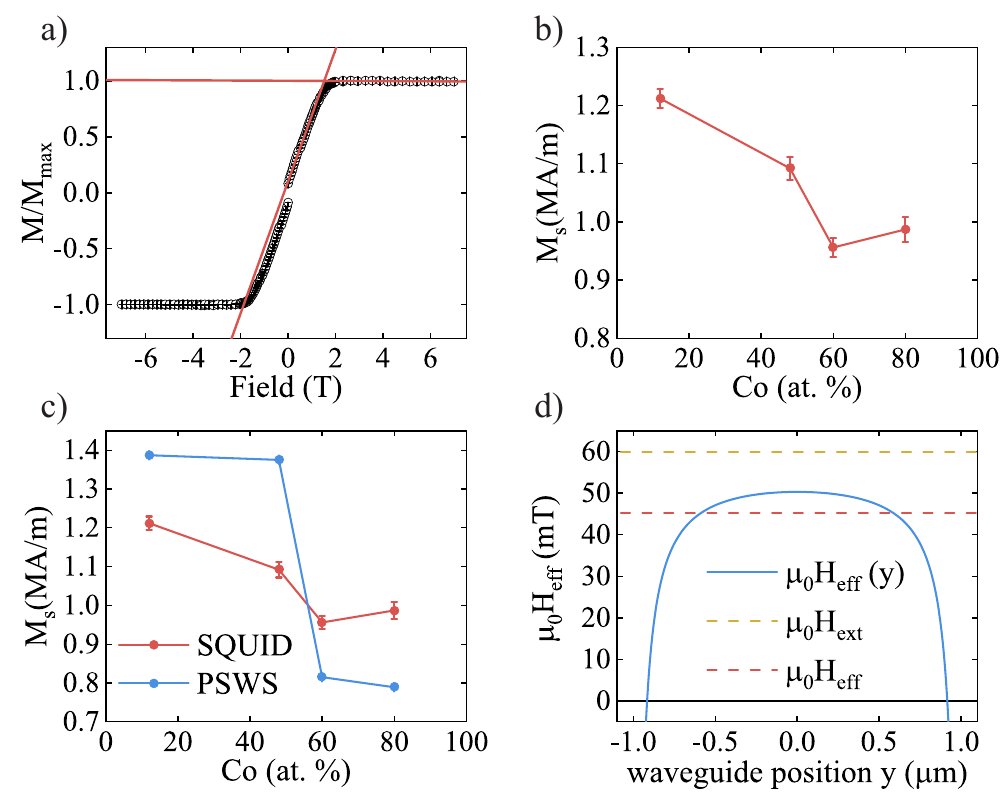}
    \caption{\label{fig4} a) Out of plain SQUID loop. b) Extracted $M_{\mathrm{s}}$ for the $\mathrm{Ta}(4)/\mathrm{Co}_{\rm x}\mathrm{Fe}_{\rm {80-x}}\mathrm{B}_{\rm 20}(20)/\mathrm{Ta}(4)$ series. c) $M_{\mathrm{s}}$ obtained with PSWS and SQUID comparison. d) Example of effective field calculated using Eq.~\ref{eq:demag} for $\mathrm{Ta(}4)/\mathrm{Co}_{\rm {12}}\mathrm{Fe}_{\rm {68}}\mathrm{B}_{\rm {20}}(20)/\mathrm{Ta}(4)$ using $M_{\rm s}$ from SQUID measurements with $H_{\rm {ext}}= \mathrm{60\ mT}$, $w=2\ \mu\mathrm{m}$, $t=20\ \mathrm{nm}$.  }
\end{figure}
where $H_{\rm {ext}}$ is the external applied magnetic field, $t$ is the thickness of the film, $w$ is the width of the strip, and $y$ is the position throughout the width with $y_{\rm 0}$ being the center. $M_{\rm s}$ was measured via superconducting quantum interference device (SQUID) by using out-of-plane hysteresis loops and extracting the saturation field as we expect negligible crystalline and interface anisotropies i.e. only shape. An example of fitting the saturation field at positive field is shown in Fig.~\ref{fig4}(a). The results for the full series are shown in Fig.~\ref{fig4}(b). 

In Fig.~\ref{fig4}(d) we plot the effective field $H_{\rm {eff}}$ along the magnetic strip width which is representative for the field of the magnetostatic surface spin wave mode (Damon Eschbach), i.e. correcting for the demagnetizing field in the plane. To obtain a representative value of this field we average $H_{\rm {eff}}$ within the region where the field is between 90\% and 100\% of its maximum value at the center of the strip $(y_{\rm 0}=0)$\cite{bracher2017creation}.

\nocite{*}
\bibliography{SUPPL_RL}